\documentclass[useAMS,usenatbib,usegraphics]{mn2e}

\usepackage[utf8]{inputenc}
\usepackage{graphicx,epsfig}  
\usepackage{amsmath,amssymb}
\usepackage{subfigure}
\usepackage{longtable}\usepackage{lscape}
\usepackage{times}
\usepackage{textcomp}
\def\ut#1{\mathop{\vtop{\ialign{##\crcr
     $\hfil\displaystyle{#1}\hfil$\crcr\noalign
     {\kern1pt\nointerlineskip}\hbox{$\hfil\sim\hfil$}\crcr
     \noalign{\kern1pt}}}}}

\def\undersymbol#1#2{\mathop{\vtop{\ialign{##\crcr
     $\hfil\displaystyle{#2}\hfil$\crcr\noalign
     {\kern1pt\nointerlineskip}\hbox{$\hfil#1\hfil$}\crcr
     \noalign{\kern1pt}}}}}

\title[Signatures of rotating binaries in micro-lensing experiments]{Signatures of rotating binaries in micro-lensing experiments}
\author[A.A. Nucita,  M. Giordano,  F. De Paolis,\& G. Ingrosso]{A.A. Nucita$^{1,2}$,  M. Giordano$^{1}$,  F. De Paolis$^{1,2}$, \& G. Ingrosso$^{1,2}$\\
$^{1}$Department of Mathematics and Physics {\it ``E. De Giorgi''}, University of Salento, Via per Arnesano, CP 193, I-73100, 
Lecce, Italy \\ 
$^{2}$INFN, Sez. di Lecce, Via per Arnesano, CP 193, I-73100, Lecce, Italy\\}  

\begin{document}

\date{Accepted xxx; Received xxx; in original form xxx}
\pagerange{\pageref{firstpage}--\pageref{lastpage}} \pubyear{2011}

\maketitle
\label{firstpage}

\begin{abstract}
Gravitational micro-lensing offers a powerful method to probe a variety of binary lens systems 
as the binarity of the lens introduces in the event light curves deviations from the typical (single lens) Paczy\'{n}ski behaviour. 
Generally, a static binary 
lens is considered to fit the observed light curve and, when
the orbital motion is taken into account, an over-simplified model is usually employed. In this paper, we treat in a realistic way the 
binary lens motion and focus on simulated events well fitted by a Paczy\'{n}ski curve. We show that, most often, an accurate timing analysis of 
the residuals (calculated with respect to the best fit Paczy\'{n}ski model) is sufficient to infer the orbital period of the binary lens. 
It goes without saying that
the independently estimated period may be used to 
further constrain the orbital parameters obtained by the best fitting procedure that often gives degenerate solutions. 
We also present a preliminary analysis on the event OGLE-2011-BLG-1127 / MOA-2011-BLG-322 which was recognized to be 
due to a binary lens. The period analysis results in 
a periodicity of $\simeq 12$ days which confirms 
the oscillation of the observed data around the best fit model. 
The estimated periodicity is likely associated to an intrinsic variability of the source star, thus opening the possibility to use this 
technique to investigate either the intrinsic variability of the source and the effects induced by the binary lens orbital motion.   
\end{abstract}


\section{Introduction}\label{sec:intro}
In the last decades the gravitational micro-lensing technique, 
initially developed to search for Massive Compact halo objects (MACHOs) 
in the Galactic halo and in the Galactic disk
\citep{pacz86,macho93,Paczynsky_96, Roulet_97,Roulet_02,Zakharov_Sazhin_UFN98}, has been widely used 
to infer the presence of exo-planets\footnote{At the time of writing, the number of exo-planets detected via 
the micro-lensing method is 24 (see e.g. the extra-solar planet encyclopedia available at \texttt{http://exoplanet.eu/}).} 
orbiting around the main lensing stars (see e.g. the reviews by \citealt{perryman2000,Perryman2005,bennett09}). 
Indeed, as shown by \cite{maopacz91} the presence of a planet around its hosting star forms a binary lens that can micro-lens 
a background source, inducing non-negligible deviations with respect 
to the usual symmetric Paczy\'{n}ski light curve (\citealt{Witt94,Gould94,Alcock97}). 
In particular, the light-curve analysis of highly magnified events is sensitive to the presence of lens companions when
the binary components are separated by a distance of the order of Einstein radius $R_{\rm E}$ associated to the whole lens system. 
In general, such signatures are characterized by short duration deviations lasting from a few hours to a few days (depending
on the parameters of the binary lens system). As a matter of fact, 
the micro-lensing technique is so sensitive that it allows to detect exo-planets in a rather large range of masses, spanning from
Jupiter-like planets down to {few Earth-mass objects} \citep{bennett96,bond04,udalski05,beaulieu06,gould06,gaudi08,bennett09,Dominik10,Gaudi10}.

Micro-lensing has also become a powerful tool to study several aspects of stellar 
astrophysics. In fact, a sufficiently amplified micro-lensing event gives the opportunity
to investigate the source limb-darkening profile, i.e. the variation of intensity from the center of 
the disc up to its border, thus implying the possibility to test among different atmosphere models (\citealt{abe2003,gaudi99}).
This opens the unique possibility to investigate the Galactic bulge star atmospheres, otherwise 
hardly to be studied due to their distance. Besides the brightness profile 
of the background source disc, highly magnified events with large radii sources allow   
to measure the lens Einstein radius once the physical radius of the source star is known from 
different methods.  
More recently, it has also been proposed (see, e.g., \citealt{ingrosso2012} and references therein) that, when finite 
size source effects become relevant,
a characteristic polarization signal may arise in micro-lensing events. This is due to the differential 
magnification induced during the crossing of the source star over the lens (either single or binary). This produces a polarization signal up to 0.04$\%$ for late type stars and up to a few percent
for cool giants, depending on the physical processes at the basis of the polarization and the atmosphere parameters of the source star. Such a signal 
may be observable with the currently available technology. It was also shown (\citealt{ingrosso2013b}) that
polarization measurements may help in disentangling the particular degeneracy between the binary or planet lens solution which occurs 
in some micro-lensing events.

Note that finite source effects usually work against the appearance of exo-planet signatures in micro-lensing light curves, 
as these features tend to be smeared out. However, although the finite source effects are generally tiny, they cannot be ignored 
when modeling a light curve. This is particularly true for large planetary masses or in case of binary micro-lensing events
characterized by caustic crossing, i.e. when the source passes through fold and/or cusp caustics  \citep{SEF}. 

The main reason for considering binary events is that the best fit procedure to a micro-lensing light curve may allow, in principle, to derive 
the parameters (the projected lens separation $b$ and the mass ratio $q$) of the lens system. Typically, a static binary lens is considered and this 
implies the minimization of a functional depending on seven free parameters. These are, in addition to $b$ and $q$, the time $t_0$ 
of closest approach to the lens system, 
the impact parameter $u_0$ (in units of the Einstein radius), the Einstein time of the event $T_E$, the angle $\theta$ that the background source trajectory forms with respect to the binary lens separation,
and the source radius $\rho_*$. Considering in the fit procedure 
the orbital motion of the lens system (see e.g. \citealt{dominik1997,penny2011a,penny2011b}) is extremely time consuming since a good modeling of such motion would imply six additional parameters, 
i.e. the semi-major axis $a$, the orbit eccentricity $e$, 
the time of passage at periastron $t_p$, the angle $i$ between the normal to lens orbital plane and 
the line of sight, the orientation $\phi_a$ of the orbital plane in the sky, and the orbiting versus (either clockwise or counterclockwise).  
As discussed in \citet{park2013}, for the determination of the lens parameters one should also include the relative lens-source parallax $\pi_E$ due to Earth motion around the Sun which 
involves two more parameters. As a result of the large number of parameters involved\footnote{Note that in the most general case one should also consider the baseline magnitude and the blending parameter.}, 
when a best fit procedure is attempted, some tricks are required in order to make the fit converging to reliable results. 

In this paper, we will concentrate on binary systems with orbital parameters for which the resulting micro-lensing light curve is very close
to a Paczy\'{n}ski curve (i.e. a planetary case). Indeed, one expects that the presence of planets rotating around the hosting star 
should cause, most often, only small perturbations (see also \citealt{bozza1999}) to the Paczy\'{n}ski light curve 
associated to an equivalent single lens event. We consider two cases: 1) the binary system orbital period $P$ is 
lower than the typical Einstein time $T_E$ of the event, and 2) $P$ is comparable or larger than $T_E$. In the latter, we 
required that the light curve is long enough to contain
at least three full cycles. Then we showed that an accurate timing analysis of the residuals (calculated with respect to the Paczy\'{n}ski model) 
is sufficient to infer $P$. 

The plan of the paper is as follows: in Section 2, we describe the model adopted for simulating a micro-lensing light curve 
taking into account the orbital motion of the binary system and the finite source size effects. In Section 3, for selected simulated events, 
we apply the timing analysis to best fit residuals and, finally, in Section 4, we discuss the advantages of our procedure and 
present a preliminary analysis on the OGLE-2011-BLG-1127 / MOA-2011-BLG-322 event, and address future perspectives.   


\section{Simulating a rotating binary micro-lensing event}
\subsection{The complex notation for a binary with point-like masses}
It is well known (see e.g. \citealt{SEF}) that in a binary micro-lensing event involving a point-like source
the angular positions of the images (${\bf x}$) and binary lens components (${\bf x_A}$ and ${\bf x_B}$)
in the lens plane are related to the angular source
position (${\bf y}$) in the source plane via the lens equation
\begin{equation}
{\bf y}={\bf x}-m_A \frac{{\bf x}-{\bf x_A}}{|{\bf x}-{\bf x_A}|^2}-m_B \frac{{\bf x}-{\bf x_B}}{|{\bf x}-{\bf x_B}|^2}~,
\label{lenseq}
\end{equation}
where $m_A$ and $m_B$ are the masses of the binary lens components {normalized to the total system mass} with $m_A>m_B$,  $q=m_B/m_A$ and $m_A+m_B=1$. 
All the positions are measured in units of the angular 
Einstein radius $\theta_E$ of the whole lens system, i.e.
\begin{equation}
\theta_E=\sqrt{\frac{4GM}{c^2}\frac{D_{LS}}{D_{OS}D_{OL}}}~,
\label{einsteinradius1}
\end{equation}
being $D_{OS}$ and $D_{OL}$ (with $D_{LS}=D_{OS}-D_{OL}$) the distance from the observer to the 
source and lens, respectively. Here, $G$ is the gravitational constant and $c$ the speed of light.
Eq. (\ref{lenseq}) can be easily generalized in order to account for multiple 
lenses as in \citet{Kayser}. 
By using complex notation (see \citealt{witt90,wittpetters1993,witt95}), eq. (\ref{lenseq}) can be easily rewritten in terms of complex variables. 
Let us define the variables $z\equiv(x,y)=x+iy$ and $\zeta\equiv(\xi,\eta)=\xi+i\eta$, being $(x,y)$ and $(\xi,\eta)$ the dimensionless 
components of the vector ${\bf x}$ and ${\bf y}$, respectively. With this substitution one has
\begin{equation}
\zeta=z-\frac{m_A}{\bar{z}-\bar{z}_A}-\frac{m_B}{\bar{z}-\bar{z}_B}~,
\label{lenseq2}
\end{equation}
where the bar over a symbol indicates the operator of complex conjugation. 
  
To determine the image position and magnification, one has to take the complex conjugate of eq. (\ref{lenseq2})
and substitute the expression for $\bar{z}$ back in it. After some algebra, one gets a fifth-order polynomial 
in $z$, i. e. $p(z)=\sum_{i=0}^{5}c_iz^i=0$ with coefficients $c_i$ depending on $b$, and $q$, whose solutions give the image positions. Here, 
$b$ represents the separation of the two binary lens components in a reference frame ($O~\xi \eta \kappa$) with origin in 
the middle point and $\xi$ axis oriented from the star to the planet and so that $\xi_A=-b/2$ and $\xi_B=+b/2$, i.e. the star and its companion are 
always on the $\xi$ (real) axis. The vertical axis $\kappa$ is oriented in order to have a right-handed reference frame. 
A (fixed) reference frame $CM~\xi'\eta'\kappa'$, with origin in the center of mass  and axes $\xi'$ and $\eta'$ parallel to $\xi$ and $\eta$,  
will be always used in the following.    

The lensing phenomenon separates the source star image into several pieces ($N_I$) and conserves the source brightness (see, e.g., \citealt{SEF}). 
Hence, for a point-like source, the summation over the magnification of each of the images gives the total magnification \citep{witt95}
\begin{equation}
A_0({\bf y}) = \sum_i^{N_I}\left(\frac{\pi_{i}}{\det J({\bf x_i})}\right)~,
\label{amplipointlike}
\end{equation}
where $\pi_{i}$ is the parity of the i-th image and the determinant of the Jacobian is
\begin{equation}
\det J = 1-\frac{\partial \xi}
{\partial \overline{z}}\frac{\overline{\partial \xi}}{\partial \overline{z}}~.
\end{equation}
The point-source magnification $A_0({\bf y})$ diverges if any of the images $\bf x_j$ appear on the critical curve of the lens, i.e. the 
curve along which $\det J$ vanishes. This curve, once mapped back in the source plane via the lens equation in eq. (\ref{lenseq}) or, equivalently, 
in eq. (\ref{lenseq2}), gives the caustic curves.
As shown by  \citet{witt90}, the critical curves can be obtained by solving 
\begin{equation}
\frac{m_A}{(\bar{z}-\bar{z}_A)^2}+\frac{m_B}{(\bar{z}-\bar{z}_B)^2}=\exp^{i\psi}~,
\label{eq6}
\end{equation}
with real phase $\psi$ varying from $0$ to $2\pi$. The previous equation corresponds to a fourth-order complex polynomial in z, the 
roots of which\footnote{In the present work we use the new 
and robust algorithm to solve polynomial 
equations presented in \citet{gould2012} and optimized for the micro-lensing case. The code is a few times faster than similar methods 
designed for more general purposes.} lie on the critical curves. By varying $\psi$ and searching for the roots of eq. (\ref{eq6}), 
one finally gets the caustic curves via the lens equation\footnote{For a binary system of point-like masses, the caustic curve is
constituted either by a single continuous curve or two (or three) 
separate paths depending on the binary parameters $b$ and $q$ (\citealt{SEF}). For a complete picture of 
the caustic zoo, we {refer to \citet{pejcha2008}.}}.

\subsection{The Inverse Ray Shooting (IRS) method}
The complex method described above fails when the source is very close to the lens 
caustics. In such conditions, a more accurate (but more time consuming) approach is given by 
the IRS technique (see e.g. \cite{SEF,Kayser,Wambsganss}). Note that the 
IRS method is a general technique, thus allowing one to consider all those 
effects which can not be easily implemented when generating a synthetic light curve. 
The method consists in solving numerically 
the lens equation in eq. (\ref{lenseq2}) by shooting light rays backwards from the observer 
through the lens plane up to the source plane. The rays which are collected in the source plane, i.e. those for which the 
{two sides of the lens equation} differ at least by a chosen tolerance, and the density of the rays at a particular location in the source plane
is proportional to the magnification at this point. Also in this case
the magnification map depends on the mass ratio $q$ and the 
projected binary lens separation $b$. As one can easily verify, the complex and IRS methods give 
comparable results beyond a distance $R$ (in Einstein radii in the source plane) from the fold and cusp caustics. Hence, we  
use the IRS method for any source distance less than $R$, and the complex method otherwise. We verified that a good compromise 
between the calculation speed and the quality of the output simulated light curve was obtained by 
choosing $R=0.05$  $R_{\rm E}$. 
\begin{figure*}
\vspace{15.cm} 
\begin{center}
$
\begin{array}{cc}
\includegraphics{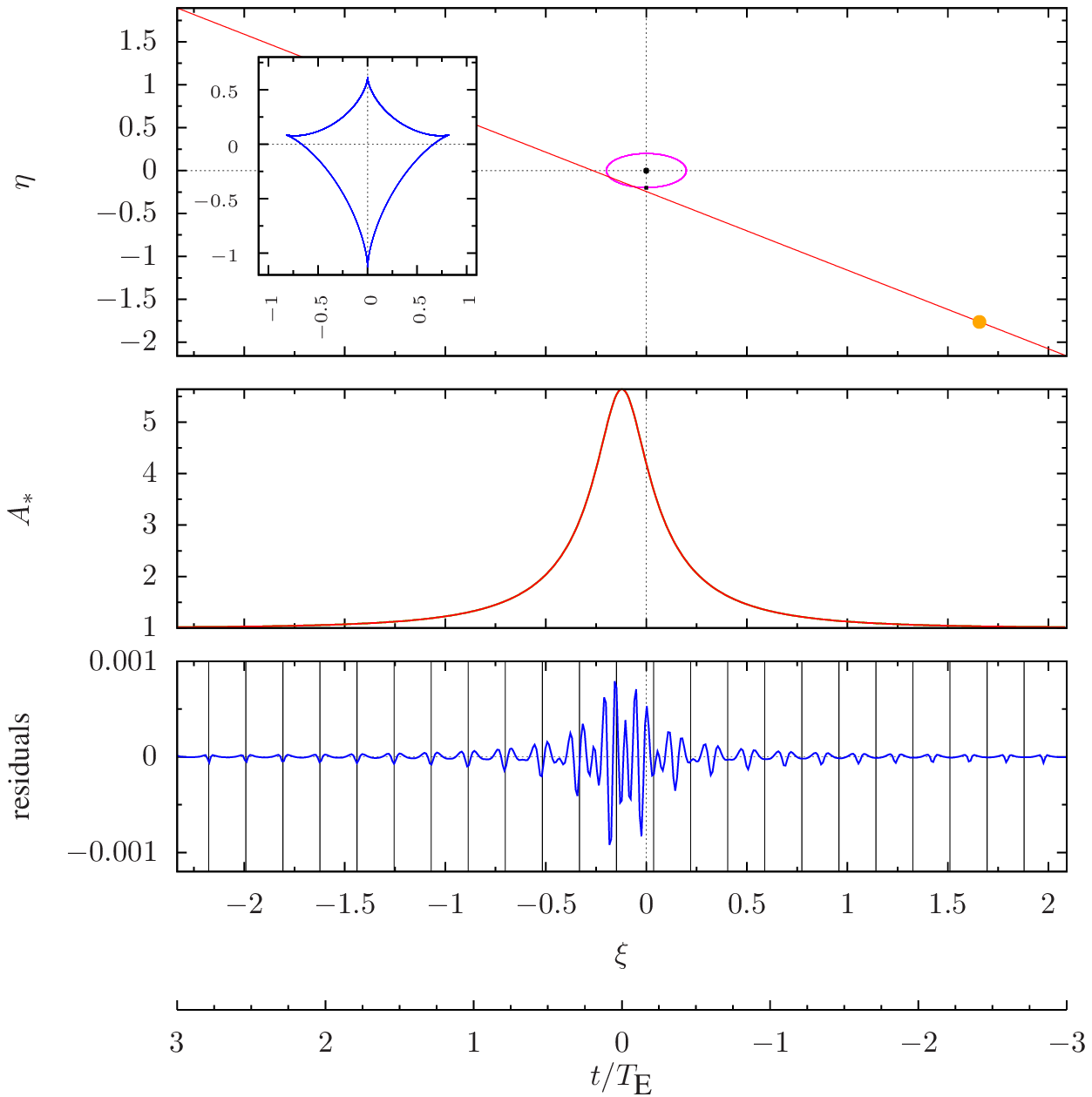} &
 \includegraphics{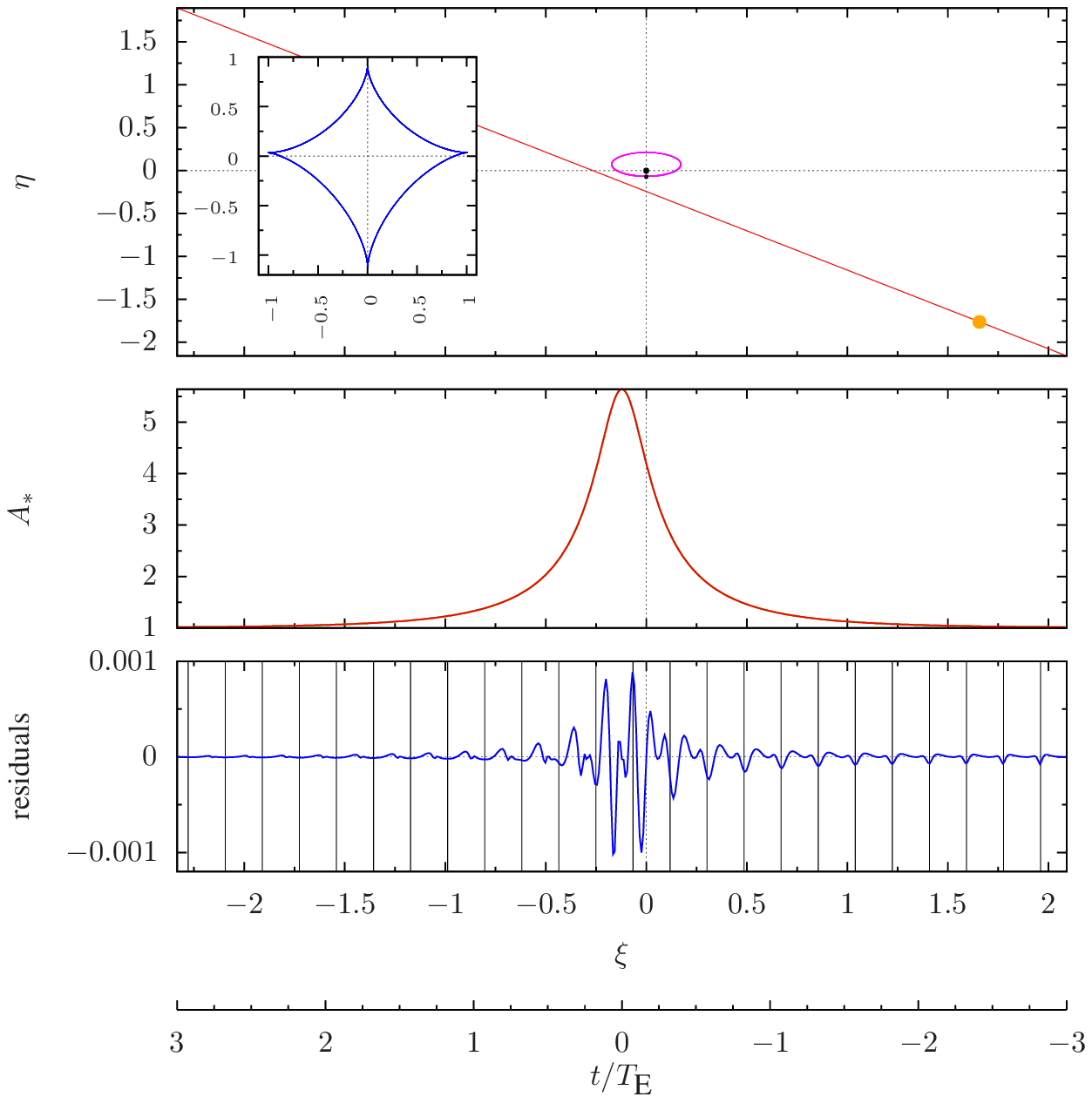}\\

\includegraphics{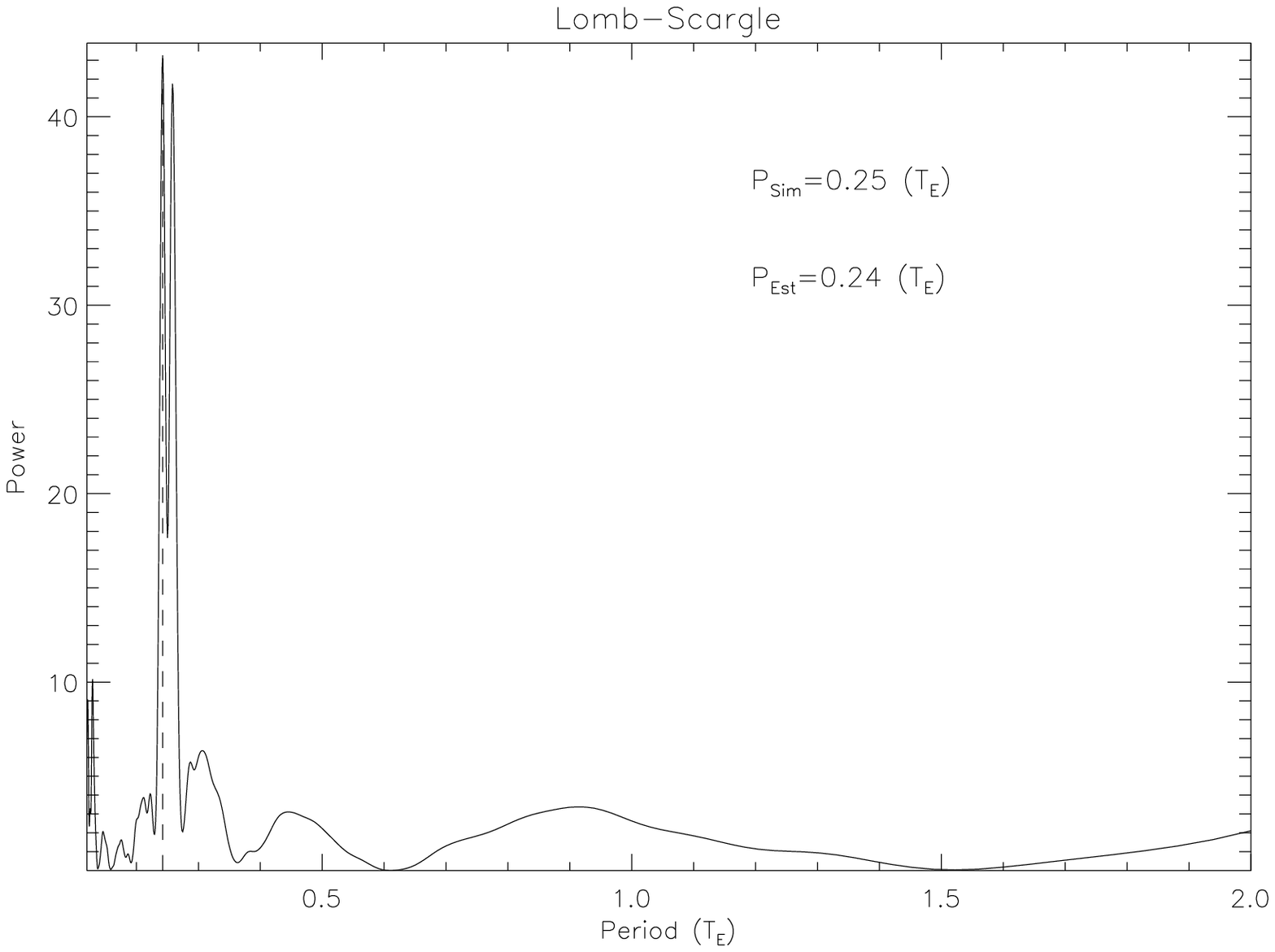} &
 \includegraphics{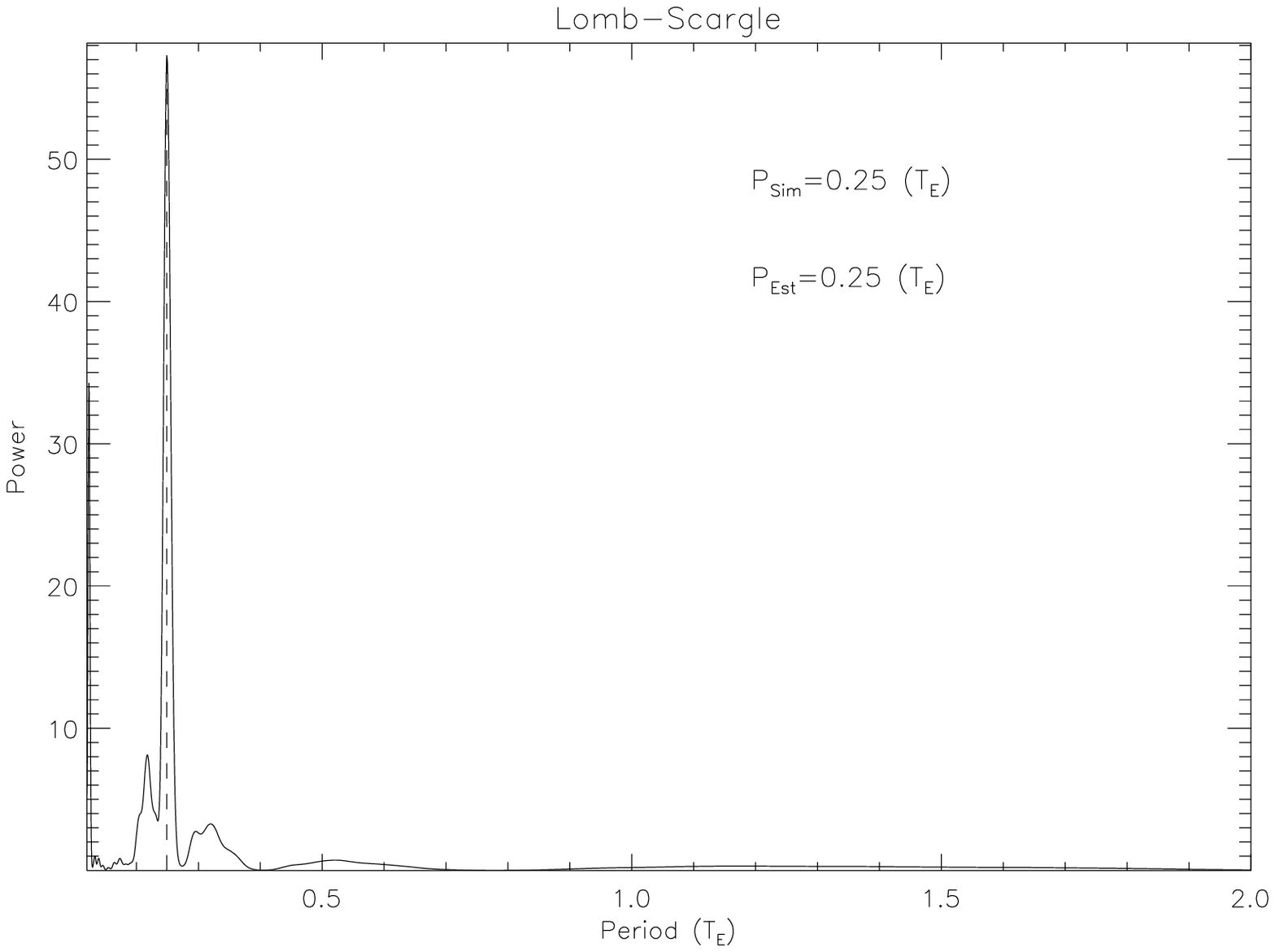} \\
\end{array}
$
\end{center}
\caption{
Left side: a snapshot at a generic time $t$ of the position of the source and of the rotating lenses is given (upper panel) together with a zoom (insets) 
around the associated central caustic curve. 
Here, the gravitational lens is assumed to be a binary system rotating counterclockwise with parameters $a=0.2$, $q=1\times 10^{-3}$, $e=0$, and $i=\phi_a=0\degr$ so that the 
orbital period $P_{Sim}=0.25 T_E$. The background source (with radius $\rho_*=0.03$) is moving with impact parameter $u_0=0.18$ (as measured from
the binary center of mass), at an angle $\theta\simeq 138\degr$ with respect to the $\xi$ axis (i.e. from right to left). In the middle panel, 
we give the magnification light curve (red line) together with 
the best fit Paczy\'{n}ski model (green line), here practically superimposed. The residuals between the simulated data and the best-fit model are also shown. In the bottom panel the 
Lomb-Scargle periodogram is given with the dashed line indicating the period ($P_{Est}$) recovered by the analysis (see discussion in the text). Right side: the same as before but for different values of the eccentricity, 
inclination and phase angles ($e=0.5$, $i=45\degr$, and $\phi_a=0\degr$) of the binary system. For clarity, the tick labels of the insets were multiplied by $10^{4}$ (left) and $10^5$ 
(right), respectively. 
}
\label{fig1}
\end{figure*}

\subsection{Finite source effects}
For real astrophysical source stars, the point-like source approximation 
described above breaks down where the magnification $A_0({\bf y})$ varies non linearly 
on the scales comparable to the source radius $\rho_*$. In this situation, the magnification 
is obtained by integrating $A_0({\bf y})$ in eq. (\ref{amplipointlike}) - weighted with the surface brightness 
$S({\bf y})$ - over the source disk and dividing by 
the unamplified source flux. In particular, 
\begin{equation}
A_*({\bf y_c}) =\frac{\int_{\Sigma_S}A_0({\bf y_c+y'})S({\bf y'}) d^2{\bf y'}}{\int_{\Sigma_S} S({\bf y'})d^2{\bf y'}}~,
\label{amplifinite}
\end{equation}
where ${\bf y_c}$ is the center of the background source star on its trajectory at a given time $t$ and the surface integral 
is extended over the whole surface $\Sigma_S=\pi \rho_*^2$.

We numerically solve\footnote{The numerical integration is performed by using the Cuhre method implemented in \citet{hahn2005}.} 
the previous integral when the source (or part of it) is close enough (or crosses) the caustic curves associated to the binary lens 
system, i.e. when the IRS method has to be employed ($R\le 0.05$  $R_{\rm E}$). 
Otherwise, for any source position for which the complex formalism \citep{witt90} is a good approximation, we use the hexadecapole technique 
(\citealt{gould2008}). The hexadecapole method substantially consists in calculating the magnification in 13 locations of the source disk:
one position coincident with the star center -the monopole approximation-, eight positions on the source limb -with radius $\rho_*$-, and four positions on 
a ring of radius $\rho_*/2$. 
As a last remark, we remind that we assumed the source surface brightness to be described by the linear limb-darkening profile \citep{Choi}
\begin{eqnarray}
S_{\lambda}(y) =  
\left[ 1- \Gamma_{\lambda} \left(1-\frac{3}{2} y \right)\right]~,  
\label{linear_LD} 
\end{eqnarray}
where the parameter $\Gamma_{\lambda}$ depends on wavelength, spectral type, surface gravity 
and metallicity of the source star. In the following, we assume $\Gamma_{\lambda} = 0.5$ as a typical value \citep{Claret}.

\subsection{The lens system orbital motion}
For a binary system constituted by two masses $m_A$ and $m_B$ orbiting around the common center of mass, the reduced mass $\mu = m_Am_B/(m_A+m_B)$ 
moves on an ellipse, with semi-major axis $a$ and 
eccentricity $e$. The orbit equation\footnote{In this work, we assume that the binary lens motion occurs with respect 
to a right-handed reference frame ($CM~XYZ$) with origin in one of the ellipse foci. 
The trajectory is in the $XY$ plane, being  the $Z$ axis orthogonal to it. In general, the orbit of the reduced
mass is also characterized by a semi-major axis forming a phase angle $\phi_a$ (measured counterclockwise) with the $X$ axis. Finally, the orbital plane is  
seen by the distant observer with an inclination angle $i$ between the $Z$ axis and the line of sight. Note that three rotations and a traslation (depending on time $t$) are required in order to switch between 
the $CM XYZ$ and $O~\xi \eta \kappa$ reference frames.} is (see e.g. \citealt{smart})
\begin{equation}
r(\psi)=a(1-e\cos \psi)~,
\label{radialreduced}
\end{equation}
where $r(\psi)$ is the radial distance from the center of mass and $\psi$ the {\it true anomaly}, respectively. 
The true anomaly is related to the {\it mean anomaly} by the Kepler equation 
\begin{equation}
\phi\equiv\omega(t-t_p)=\psi-e\sin \psi~,
\end{equation}
being $t$ a generic instant of time during the orbital motion, $t_p$ the time of the pericenter passage and $\omega$ the angular frequency given, as usual, by 

\begin{equation}
\omega\equiv \frac{2\pi}{P}=\sqrt{\frac{GM}{a^3}}~,
\label{keplerperiod}
\end{equation}
where $P$ is the keplerian orbital period. Since we are giving the semi-major axis $a$ in units of the Einstein radius, 
the orbital period $P$ results to be given in Einstein time. 
\begin{figure*}
\vspace{15.cm} 
\begin{center}
$
\begin{array}{cc}
\includegraphics{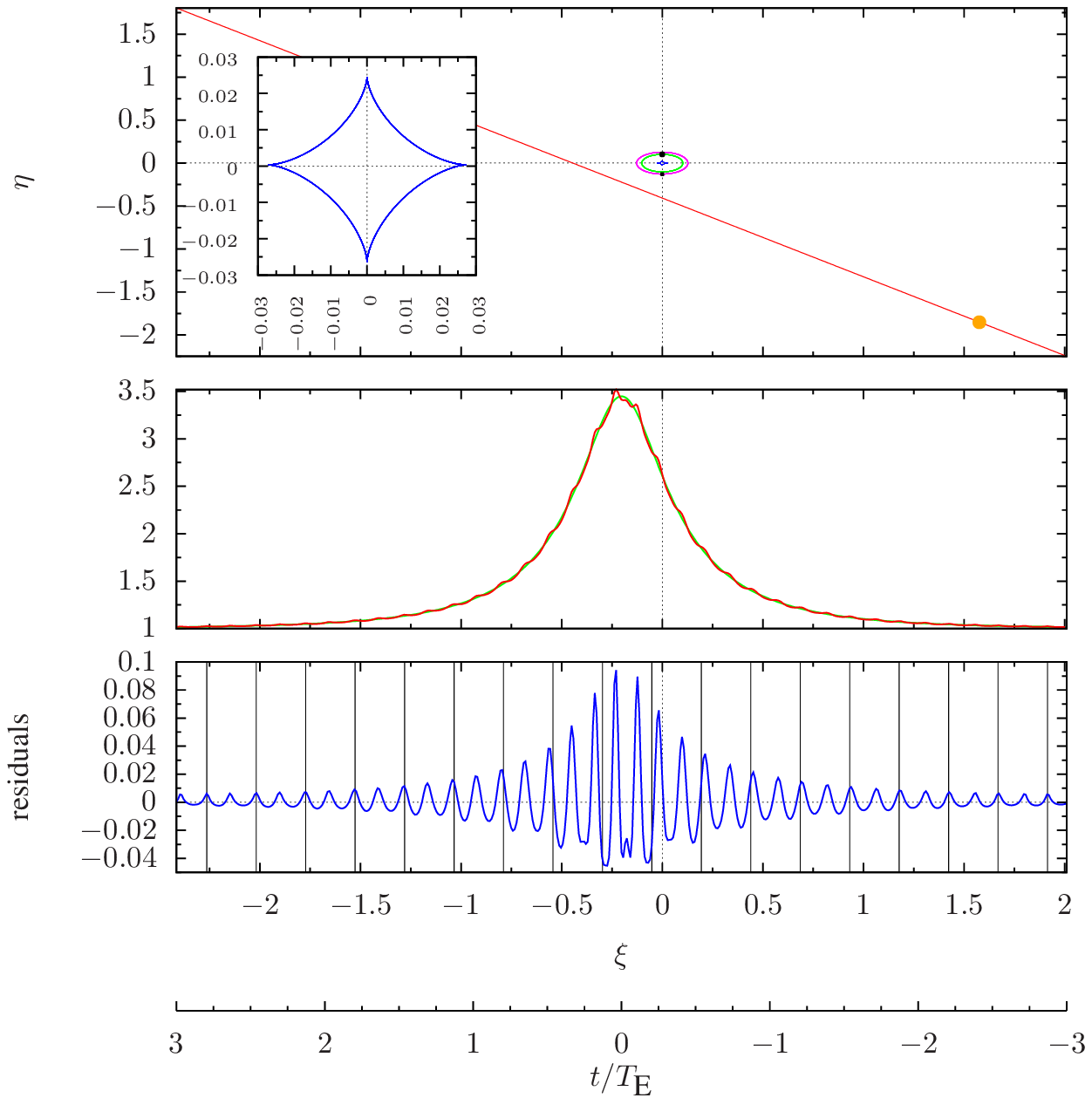} &
 \includegraphics{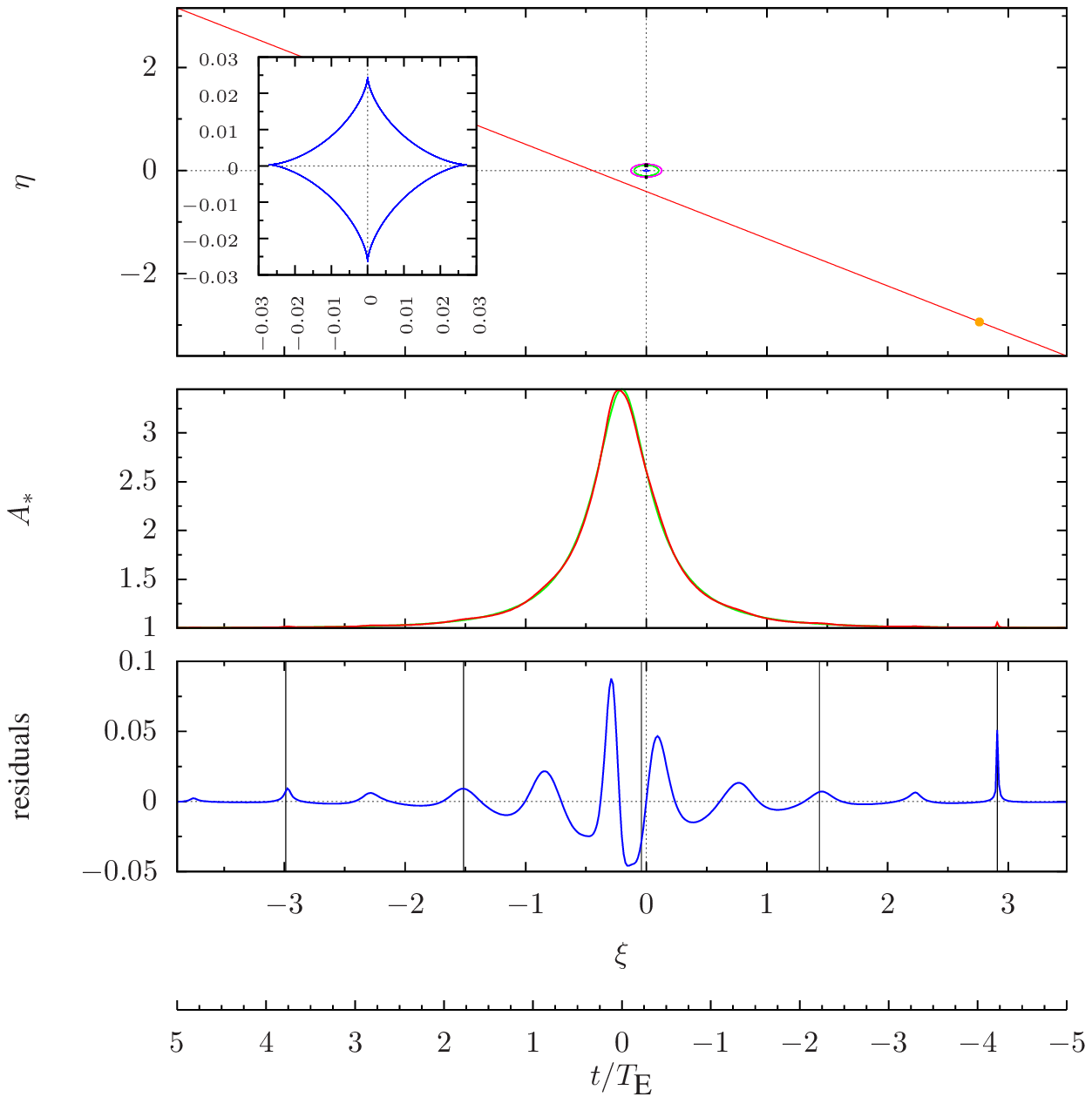}\\

\includegraphics{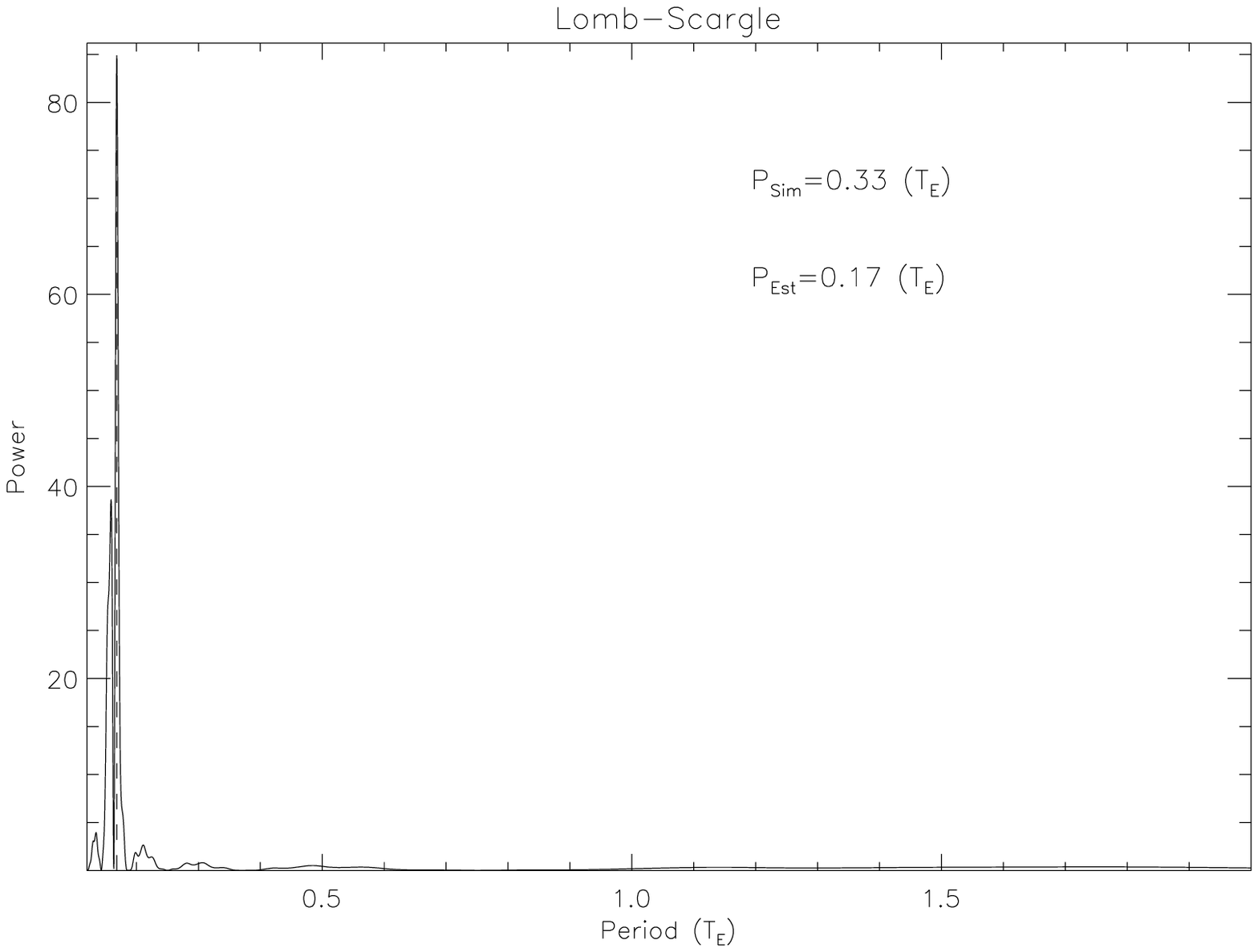} &
 \includegraphics{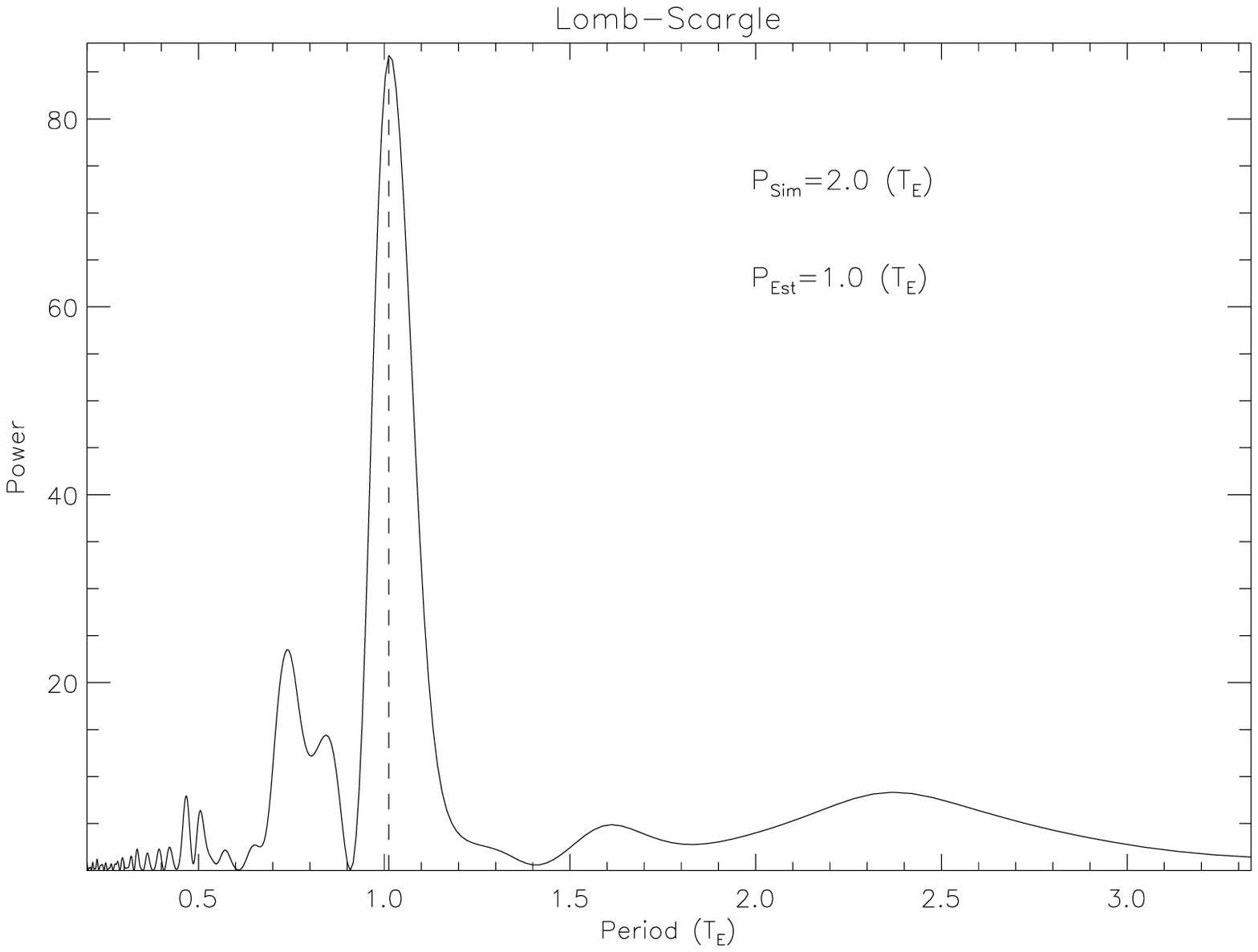} \\
\end{array}
$
\end{center}
\caption{Left side: micro-lensing light curve for a binary system with $a=0.23$, $q=0.8$, $e=0$ and $i=\phi_a=0\degr$. 
The source and trajectory parameters are the same as in the case described in Fig. 1. The system is characterized 
by an orbital period $P_{Sim}\simeq 0.33 T_E$, and the event lasts for $\simeq 6T_E$. Right side: by using the same orbital parameters, 
we fixed the orbital period to $\simeq 2 T_E$ and an observation time of $\simeq 10 T_E$.
The large (short duration) spike clearly visible in the residual light curve is due to a planetary caustic crossing event which occurred during the source transit. The bottom panels show the Lomb-Scargle 
periodograms as discussed in the text.} 
\label{fig2}
\end{figure*}

We then solved the Kepler equation by using a series expansion of Bessel functions (see e.g. \citealt{watson}). In particular, it can be easily shown that
\begin{equation}
\psi=\phi+\sum_{n=1}^{+\infty}\frac{2}{n}J_n(ne)\sin (n\phi)~,
\label{besselsolution}
\end{equation}
where $J_n(ne)$ is the Bessel function of order $n$, and the mean anomaly $\phi$ contains the time dependence. Note that a few terms in the previous 
summation are usually required to get the solution of the Kepler equation within a few percent of accuracy. 
Hence, for any time $t$, the previous 
equation gives the true anomaly $\psi$ with 
the reduced mass position obtained by means of eq. (\ref{radialreduced}). With this formalism, the radial coordinates of the two lenses on their orbits are 
\begin{equation}
r_A(\psi)=-r(\psi)\mu/m_A ~,~~~~~r_B(\psi)=r(\psi)\mu/m_B ~.
\end{equation}
Of course, the separation $b$ (which enters in the complex and IRS methods described previously) between the two 
lens components is now time dependent and it is 
given by the projection onto the lens plane of the vector ${\bf r_A}-{\bf r_B}$.

\section{The output synthetic light curve and the timing analysis on the residuals}
The formalism introduced in the previous Section allows us to obtain 
simulated micro-lensing light curves taking into account the orbital motion of the binary lens. Here, we concentrate on binary systems characterized by 
orbital periods $P$ lower than the typical Einstein time $T_E$ of the event and by orbital parameters $b$ and $q$ 
giving rise to event light curves close to the Paczy\'{n}ski behaviour. 

In Fig. \ref{fig1} (left side), we give a snapshot at a generic time $t$ of the position of the source, the 
rotating lens system and the associated caustic curves (inset) 
in the source plane (upper panel). In the additional material available on-line, we present a movie of the micro-lensing event corresponding to 
Fig. \ref{fig1}.  By measuring all the distances in units of the Einstein radius, we simulated a binary system rotating counterclockwise with parameters $a=0.2$, 
$q=1\times 10^{-3}$, $e=0$, and $i=\phi_a=0\degr$. The corresponding binary system has an orbital period $P_{Sim}=0.25 T_E$, being $T_E$ the Einstein time of the event. 
The purple ellipse represents the planet trajectory as projected in the plane of the sky. The background source (with radius $\rho_*=0.03$) 
is moving at the impact parameter $u_0=0.2$ (as measured from the binary center of mass)  at an 
angle $\theta\simeq 138\degr$ with respect to the $\xi$ axis. With the assumed binary system and micro-lensing parameters, 
we expect to have only minor deviations (a planetary case) with respect to the 
single lens light curve. Indeed, in the middle panel, we show the magnification light curve (red line) 
resulting from eq. (\ref{amplifinite}) to which the best fit light curve (green line) obtained with a Paczy\'{n}ski model is superimposed. 
As one can see, the two curves are almost indistinguishable in this case.
As it is clear from the bottom panel of Fig. \ref{fig1}, the residuals between the simulated light curve and the best fit model 
are as low as a few parts per thousand but clearly show a periodic behaviour (here, for simplicity, we are ignoring the effect of any noise in the data strain).
As a guide for the eye, we also show a series of vertical solid lines separated in time by one orbital period $P$. In the right side of the figure, we simulated a micro-lensing event 
due to a binary system with the same mass and semi-major axis as before, but different values of eccentricity, inclination and phase angle ($e=0.5$, $i=45\degr$, and $\phi_a=0\degr$).  

Before discussing the results obtained with the period analysis on the residuals, we note that, due to the fact that the simulated light curve and the best-fit Paczy\'{n}ski model 
do not have the peak at the same time (and this is a general 
characteristic of such events), the residuals show an asymmetry with respect to $t_0$ and a number of extra peaks (especially concentrated around the event maximum) 
that are not correlated to the real binary system period. As we will discuss later, this introduces spurious peaks in the period analysis that might make impossible to recover the simulated period. 
Indeed, any algorithm used for the period search looks for features in phase and fails around the micro-lensing maximum to give the right period. 
\begin{figure}
\vspace{15.cm} 
\begin{center}
$
\begin{array}{cc}
\includegraphics{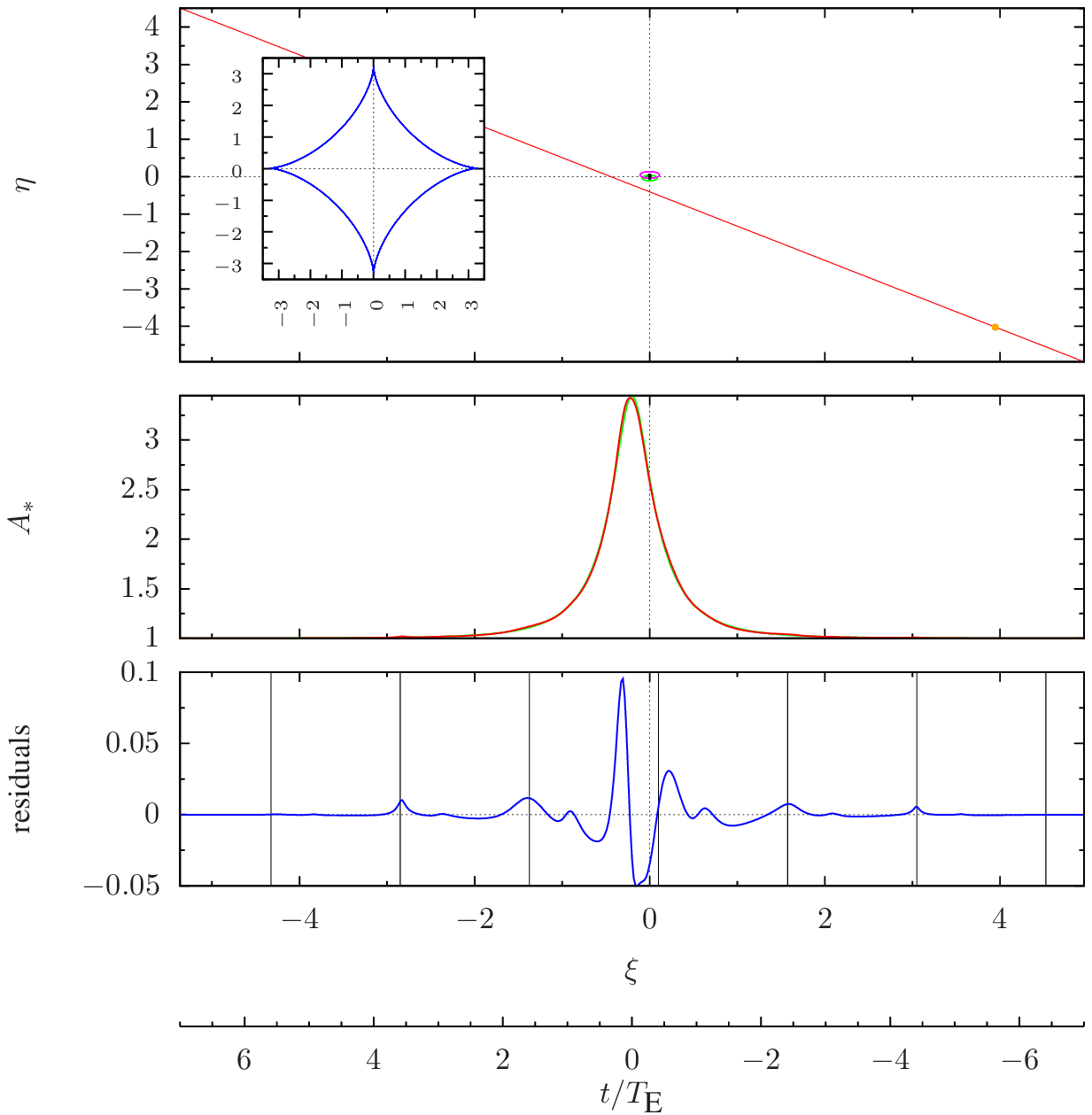}\\

\includegraphics{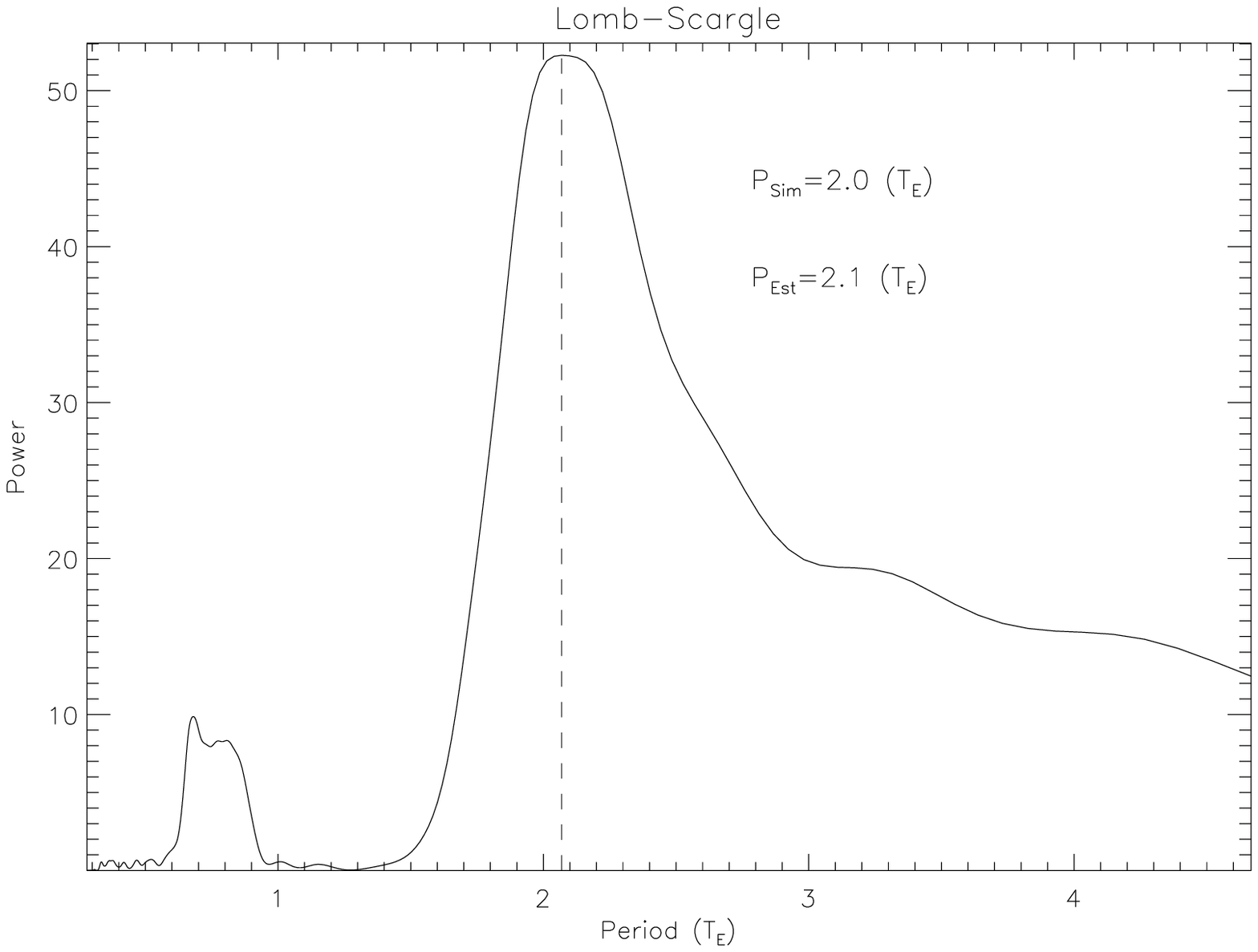}\\
\end{array}
$
\caption{The expected micro-lensing light curve for a binary system with orbital parameters $a=0.23$, 
$q=0.8$, $e=0.5$, $i=45\degr$, and $\phi_a=0\degr$, assuming the orbital period $P_{Sim}\simeq 2 T_E$ and a total observation duration of $14 T_E$. For clarity, the tick labels of the inset were multiplied by $10^{3}$. 
}
\end{center}
\label{fig3}
\end{figure}

We tested different methods to extract the period from the residual light curve for several simulated events:
the Fast Fourier Transform -FFT- \citep{nr2007}; the Phase Dispersion Minimization -PDM- \citep{stellingwerf1978}; the Lomb-Scargle periodogram \citep{lomb,scargle}; 
its generalized form given in \citet{zk2009} and the epoch folding technique. All the algorithms give consistent results, but here we prefer to present those obtained via the classical Lomb-Scargle method
(the generalized version did not improved the period detection) because its statistical behaviour is well known and the technique can be easily implemented. 

The method requires to specify the minimum ($\nu_{min}$) and maximum ($\nu_{max}$) frequencies to be searched for in the input signal, together with the frequency step 
$\Delta \nu$. We chose to set these parameters depending on the {\it observational} data set, i.e. $\nu_{min}=1/(3 T_{obs})$,  $\nu_{max}=1/(2 \delta t)$, oversampling by a factor 10, being $T_{obs}$ the duration 
of the observation and $\delta t$ the associated time step. Note that by using the minimum frequency $\nu_{min}$ we are implicitly requiring to have at least three full cycles per observational window.

The blind application of the Lomb-Scargle method to the residual light curves resulted in the periodograms shown in Fig. \ref{fig1}, where the dashed vertical lines (always at period 
$P_{Sim}\simeq 2000$) mark the 
periodicity detected in the signal. Note that the Lomb-Scargle periodogram (as well as the other period search algorithms) does not give exactly the simulated period since the frequency of the periodic signal is affected 
by the relative source-lens motion. As discussed above, the central part of the residual light curve has spurious peaks that disturb the timing analysis since they introduce additional power at frequencies 
different from the fundamental one, i.e $\nu=2\pi/P_{Sim}$.


We have found that the best results in the period search are obtained by removing a central region in the residual curve, around the event peak. Without this cut, the Lomb-Scargle periodogram may return 
other peaks in addition to that corresponding to the simulated one. It is remarkable that, in all the cases we have considered, it is sufficient to remove a 
very small region around the maximum magnification with size of the order of a fraction of the orbital period $P_{Sim}$ in order to get the true periodicity 
(indicated as $P_{Est}$).

The second case we discuss is a rotating binary similar to that considered in \citet{penny2011b} with orbital parameters $a=0.23$, $q=0.8$, $e=0$, and $i=\phi_a=0\degr$ rotating with a 
period $P_{Sim}\simeq 0.33T_E$. Since the system is face-on and the eccentricity is null, the projected distance does not change during the 
micro-lensing event, thus implying that 
the caustics simply rotate without any deformation. This particular geometry appears in the periodogram as a peak always at half of the simulated period 
(see the left bottom panel of Fig.\ref{fig2}) as a consequence of the North-South symmetry in the caustic plane. In the right side of the same figure, we simulate a binary lens event 
with the same orbital parameters but with $P_{Sim}\simeq 2 T_E$ and an observational window of $\simeq 6T_E$. 
In spite of the fact that there are only a few full cycles, the periodogram gives again half of the expected period.

We note that the geometry of the binary lens described in the two examples of Fig. \ref{fig2} is rather unlikely to appear in real observations and so, in Fig. \ref{fig3}, we consider a more realistic 
binary configuration with $e=0.5$ and $i=45\degr$, being all the other parameters unchanged. As one can see, in this case the Lomb-Scargle analysis allows us to recover the right period even if the associated periodogram peak is 
quiet broad. 

\section{Result and discussions}
\label{results3} 
\begin{figure*}
\vspace{11.0cm} \includegraphics{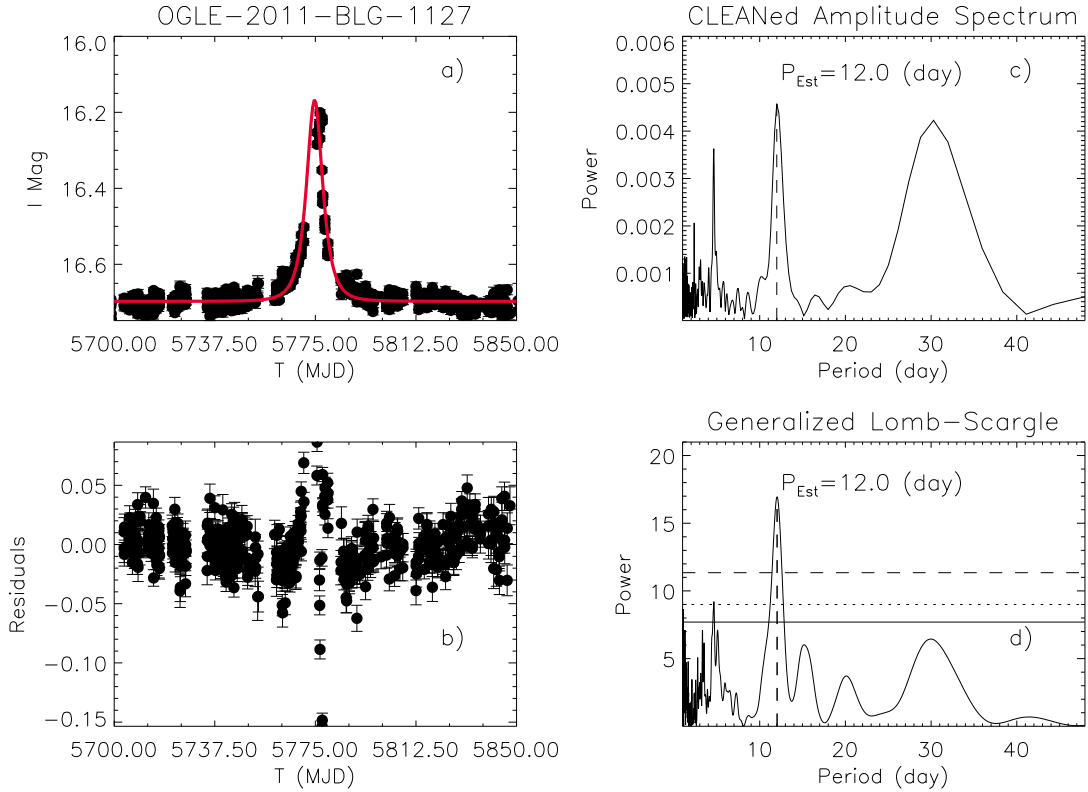}
\caption{The event OGLE-2011-BLG-1127 light curve in the I band and the best fit Paczy\'{n}ski model is shown in panel a), together with the associated residuals (panel b). In panels c) and d), we show 
the periodograms obtained via the CLEANed DFFT and generalized Lomb-Scargle methods, respectively (see text for details). For the Lomb-Scargle periodogram, we evaluated
the false alarm probability and obtained the confidence levels at $68\%$, $90\%$, and $99\%$ shown in the right-bottom panel as the solid, dotted and dashed horizontal lines, respectively.
}
\label{figogle}
\end{figure*}
We have considered the orbital binary lens motion in simulated 
micro-lensing events that have light-curves similar to the 
Paczy\'{n}ski behaviour and have shown that in most cases a timing 
analysis of the residuals allows one to recover the binary period $P$. 
Since the simulated light curve and the best-fit Paczy\'{n}ski model 
do not have, in general, the peak at the same time, the residuals show an asymmetry with respect to $t_0$ and a number of extra peaks that are not correlated 
to the real binary system period. This introduces spurious features in the analysis that can be overcome by excluding a central 
region of the light  curve whose size is reduced smoothly until the true period is lost. 

The importance of this procedure is that of inferring the orbital period of the binary lens without performing 
a numerically heavy multi-dimensional fit procedure which, generally, do not give unique solutions. Here, by considering  Paczy\'{n}ski-like 
events, we showed that the procedure is robust enough to recover the simulated orbital period, provided that the event duration contains at least 
three full cycles.

A systematic search for such signatures on real (archival) observed micro-lensing events will be presented elsewhere. Here, we show a preliminary analysis 
on the event OGLE-2011-BLG-1127 / MOA-2011-BLG-322 (\citealt{moa2011}) possibly due to a binary lens.  
By using the publicly available OGLE data (\citealt{udalski03}, http://ogle.astrouw.edu.pl/) we have fitted  the data with a simple Paczy\'{n}ski model (see Fig. \ref{figogle}, panel a) 
and analyzed the residuals (Fig. \ref{figogle}, panel b) by using either the generalized Lomb-Scargle and the discrete fast Fourier transform techniques (DFFT treated via the CLEAN algorithm of \citealt{roberts1987}). 
These are the best-performing period search methods for this kind of data. Both techniques return the presence of a period at about $\simeq 12$ days (see Fig. \ref{figogle}, panels c and d)\footnote{Here we note that the 
region around the event peak has not been removed since already poor of data.}.

It seems that this periodic feature is remarkably stable since it always appears when a time resolved periodogram is performed, i.e. when the period search algorithm is applied on different parts of the residual light curve, provided 
that each part contains a sufficiently large number of cycles. It is therefore possible that the periodicity at $\simeq 12$ days is associated to an intrinsic variability of the source star as if it is in 
a binary system or intrinsically variable ({see e.g. \citealt{wyrzy2006}}) or contaminated by a close variable star. {Note also that some period search techniques, such as the cleaned DFFT (Fig. \ref{figogle}, panels c) 
and the generalized Lomb-Scargle periodogram, give significant power also at $\simeq 4$ days and $\simeq 30$ days. Since we test trial periods in a given range,
it is possible to evaluate the significance of each feature compared to the power at all other frequencies. Hence, following \citet{hornebaliunas} (see also \citealt{zk2009}), 
we evaluate the false alarm probability and obtained the significance levels at $68\%$, $90\%$, and $99\%$ which are shown at the bottom-right panel of Fig. \ref{figogle} as solid, dotted and dashed lines, respectively. As one can note,
the periods at $\simeq 4$ days and $\simeq 30$ days are not significant.

Note also that interpreting the $\simeq 30$ days feature as a true periodicity could be problematic since it is comparable to the event Einstein time $T_E\simeq 23.4$ days. In addition, a time 
resolved period analysis does not return a stable result.      
As a matter of fact, we note that the light curve of the event OGLE-2011-BLG-1127 shows an oscillation around the best fit (static) model, 
as it is apparent in Fig. 1 of \citet{moa2011}, with a time scale of $\simeq 15$ days.  If the bump-like structure observed 
$\simeq 50$ days after the event peak is really present along the whole light curve (as also mentioned in \citealt{moa2011}) 
then our analysis is returning a periodicity related to the source or to the binary lens motion.}
 
%
%


\end{document}